\documentclass[12pt,a4paper]{article}
\usepackage[mathlines]{lineno}
\usepackage{graphicx}
\usepackage{xspace}
\usepackage{color}
\usepackage{colortbl}
\usepackage{amsmath}
\usepackage{ifthen}

\newboolean{pdflatex}
\setboolean{pdflatex}{false}

\newboolean{uprightparticles}
\setboolean{uprightparticles}{false}
\usepackage{amssymb}
\usepackage{amsfonts}
\usepackage{upgreek}

\usepackage[numbers, comma, square, sort&compress]{natbib}
\usepackage{hyperref}
\usepackage[all]{hypcap}

\ifthenelse{\boolean{uprightparticles}}
{

 \def\PDelta      {\ensuremath{\Delta}\xspace}                 
 \def\PXi      {\ensuremath{\Xi}\xspace}                 
 \def\PLambda      {\ensuremath{\Lambda}\xspace}                 
 \def\PSigma      {\ensuremath{\Sigma}\xspace}                 
 \def\POmega      {\ensuremath{\Omega}\xspace}                 
 \def\PUpsilon      {\ensuremath{\Upsilon}\xspace}

 \def\PB      {\ensuremath{\mathrm{B}}\xspace}                 
                  
 \def\PD      {\ensuremath{\mathrm{D}}\xspace}

 \def\PK      {\ensuremath{\mathrm{K}}\xspace}

 \def\Pc      {\ensuremath{\mathrm{c}}\xspace}

 \def\Pi      {\ensuremath{\mathrm{i}}\xspace}

}
{

 \mathchardef\PDelta="7101
 \mathchardef\PXi="7104
 \mathchardef\PLambda="7103
 \mathchardef\PSigma="7106
 \mathchardef\POmega="710A
 \mathchardef\PUpsilon="7107
                  
 \def\PB      {\ensuremath{B}\xspace}                 
                  
 \def\PD      {\ensuremath{D}\xspace}

 \def\PK      {\ensuremath{K}\xspace}

 \def\Pc      {\ensuremath{c}\xspace}

 \def\Pi      {\ensuremath{i}\xspace}

}

\def\c     {\ensuremath{\Pc}\xspace}

\def\kaon  {\ensuremath{\PK}\xspace}
\def\Kbar  {\kern 0.2em\overline{\kern -0.2em \PK}{}\xspace}

\def\Kz    {\ensuremath{\kaon^0}\xspace}
\def\Kzb   {\ensuremath{\Kbar^0}\xspace}
\def\KzKzb {\ensuremath{\Kz \kern -0.16em \Kzb}\xspace}
\def\Kp    {\ensuremath{\kaon^+}\xspace}
\def\Km    {\ensuremath{\kaon^-}\xspace}

\def\KpKm  {\ensuremath{\Kp \kern -0.16em \Km}\xspace}

\def\Dbar    {\kern 0.2em\overline{\kern -0.2em \PD}{}\xspace}
\def\D       {\ensuremath{\PD}\xspace}

\def\Dz      {\ensuremath{\D^0}\xspace}
\def\Dzb     {\ensuremath{\Dbar^0}\xspace}
\def\DzDzb   {\ensuremath{\Dz {\kern -0.16em \Dzb}}\xspace}
\def\Dp      {\ensuremath{\D^+}\xspace}
\def\Dm      {\ensuremath{\D^-}\xspace}

\def\DpDm    {\ensuremath{\Dp {\kern -0.16em \Dm}}\xspace}

\def\Bbar    {\kern 0.18em\overline{\kern -0.18em \PB}{}\xspace}

\def\Y#1S{\ensuremath{\PUpsilon{(#1S)}}\xspace}

\newcommand{\tev}{\ensuremath{\mathrm{\,Te\kern -0.1em V}}\xspace}
\newcommand{\gev}{\ensuremath{\mathrm{\,Ge\kern -0.1em V}}\xspace}
\newcommand{\mev}{\ensuremath{\mathrm{\,Me\kern -0.1em V}}\xspace}
\newcommand{\kev}{\ensuremath{\mathrm{\,ke\kern -0.1em V}}\xspace}
\newcommand{\ev}{\ensuremath{\mathrm{\,e\kern -0.1em V}}\xspace}
\newcommand{\gevc}{\ensuremath{{\mathrm{\,Ge\kern -0.1em V\!/}c}}\xspace}
\newcommand{\mevc}{\ensuremath{{\mathrm{\,Me\kern -0.1em V\!/}c}}\xspace}
\newcommand{\gevcc}{\ensuremath{{\mathrm{\,Ge\kern -0.1em V\!/}c^2}}\xspace}
\newcommand{\gevgevcccc}{\ensuremath{{\mathrm{\,Ge\kern -0.1em V^2\!/}c^4}}\xspace}
\newcommand{\mevcc}{\ensuremath{{\mathrm{\,Me\kern -0.1em V\!/}c^2}}\xspace}

\def\mbarn{\ensuremath{\rm \,mb}\xspace}
\def\mub{\ensuremath{\rm \,\upmu b}\xspace}

\def\invnb   {\ensuremath{\mbox{\,nb}^{-1}}\xspace}

\def\BR         {{\ensuremath{\cal B}\xspace}}

\def\to                 {\ensuremath{\rightarrow}\xspace}

\def\pt         {\mbox{$p_T$}\xspace}

\newcommand{\stat}{\ensuremath{\mathrm{(stat)}}\xspace}

\def\gsim{{~\raise.15em\hbox{$>$}\kern-.85em
          \lower.35em\hbox{$\sim$}~}\xspace}
\def\lsim{{~\raise.15em\hbox{$<$}\kern-.85em
          \lower.35em\hbox{$\sim$}~}\xspace}

\def\evtgen     {\mbox{\textsc{EvtGen}}\xspace}
\def\pythia     {\mbox{\textsc{Pythia}}\xspace}

\newcommand{\uncorrsyst}{\ensuremath{\mathrm{(syst)}}\xspace}
\newcommand{\corrsyst}{\ensuremath{\mathrm{(scale)}}\xspace}

\newcommand{\lumitot}{14.7\invnb}
\newcommand{\phiKK}{\ensuremath{\phi\rightarrow K^+ K^-}\xspace}
\newcommand{\KK}{\ensuremath{K^+ K^-}\xspace}
\newcommand{\tagNprobe}{tag-and-probe\xspace}

\newcommand{\eff}{\ensuremath{\varepsilon}\xspace}
\def\calL{{\ensuremath{\cal L}}\xspace}
\newcommand{\y}{\ensuremath{y}\xspace}

\def\pp{{\ensuremath{p{\kern-0.05em}p}}\xspace}

\newcommand{\pythiasix}{\pythia 6.4\xspace}

\newcommand{\XSecPtRange}{0.6 < p_T < 5.0\gevc}
\newcommand{\XSecYRange}{2.44 < y < 4.06}
\newcommand{\XSecRange}{\XSecPtRange \text{ and } \XSecYRange}
\newcommand{\XSectionProjTextSize}{}

\newcommand{\XSecResultTotalcombined}{\sigma(pp\to\phi X) = 1758 \pm 19\stat\, ^{+43}_{-14}\uncorrsyst \pm 182\corrsyst \mub}

\newcommand{\XSecResultTotalValRelLHCbMCcombined}{1.43 \pm 0.15}
\newcommand{\XSecResultTotalValRelPerugiacombined}{2.06 \pm 0.22}

\newcommand{\XSecTableBinnedcombinedA}{
\center
\XSectionProjTextSize
\small{
\begin{tabular}{cccc}
\hline\hline
\pt / \y & 2.44-2.62 & 2.62-2.80 & 2.80-2.98 \\\hline

0.6-0.8 & $1.001 \pm 0.140 \,^{+0.076}_{-0.026}$ & $0.853 \pm 0.114 \,^{+0.081}_{-0.022}$ & $1.069 \pm 0.108 \,^{+0.093}_{-0.027}$\\
0.8-1.0 & $0.959 \pm 0.112 \,^{+0.129}_{-0.015}$ & $0.797 \pm 0.084 \,^{+0.074}_{-0.012}$ & $0.819 \pm 0.079 \,^{+0.053}_{-0.012}$\\
1.0-1.2 & $0.758 \pm 0.043 \,^{+0.089}_{-0.009}$ & $0.776 \pm 0.038 \,^{+0.063}_{-0.009}$ & $0.795 \pm 0.026 \,^{+0.042}_{-0.009}$\\
1.2-1.4 & $0.648 \pm 0.033 \,^{+0.067}_{-0.009}$ & $0.627 \pm 0.028 \,^{+0.049}_{-0.008}$ & $0.604 \pm 0.026 \,^{+0.024}_{-0.008}$\\
1.4-1.6 & $0.469 \pm 0.023 \,^{+0.037}_{-0.008}$ & $0.511 \pm 0.022 \,^{+0.033}_{-0.008}$ & $0.521 \pm 0.022 \,^{+0.023}_{-0.008}$\\
1.6-1.8 & $0.422 \pm 0.020 \,^{+0.039}_{-0.008}$ & $0.381 \pm 0.017 \,^{+0.021}_{-0.007}$ & $0.409 \pm 0.018 \,^{+0.015}_{-0.007}$\\
1.8-2.0 & $0.334 \pm 0.016 \,^{+0.027}_{-0.007}$ & $0.323 \pm 0.015 \,^{+0.014}_{-0.007}$ & $0.276 \pm 0.012 \,^{+0.009}_{-0.005}$\\
2.0-2.4 & $0.209 \pm 0.008 \,^{+0.010}_{-0.004}$ & $0.192 \pm 0.007 \,^{+0.006}_{-0.003}$ & $0.201 \pm 0.007 \,^{+0.003}_{-0.003}$\\
2.4-2.8 & $0.127 \pm 0.005 \,^{+0.003}_{-0.003}$ & $0.112 \pm 0.005 \,^{+0.002}_{-0.003}$ & $0.111 \pm 0.004 \,^{+0.002}_{-0.002}$\\
2.8-3.2 & $0.078 \pm 0.004 \,^{+0.002}_{-0.002}$ & $0.069 \pm 0.003 \,^{+0.002}_{-0.002}$ & $0.063 \pm 0.003 \,^{+0.002}_{-0.002}$\\
3.2-4.0 & $0.040 \pm 0.002 \,^{+0.001}_{-0.001}$ & $0.038 \pm 0.002 \,^{+0.001}_{-0.001}$ & $0.034 \pm 0.001 \,^{+0.001}_{-0.001}$\\
4.0-5.0 & $0.014 \pm 0.001 \,^{+0.001}_{-0.001}$ & $0.014 \pm 0.001 \,^{+0.001}_{-0.000}$ & $0.011 \pm 0.001 \,^{+0.000}_{-0.000}$\\
\end{tabular} }
\vspace{-0.29cm}
}

\newcommand{\XSecTableBinnedcombinedB}{
 \center
\XSectionProjTextSize
\small{
\begin{tabular}{cccc}
\hline
\pt / \y & 2.98-3.16 & 3.16-3.34 & 3.34-3.52\\\hline
0.6-0.8 & $1.171 \pm 0.100 \,^{+0.058}_{-0.029}$ & $1.060 \pm 0.092 \,^{+0.027}_{-0.043}$ & $1.131 \pm 0.146 \,^{+0.029}_{-0.176}$\\
0.8-1.0 & $1.032 \pm 0.080 \,^{+0.049}_{-0.015}$ & $0.862 \pm 0.080 \,^{+0.014}_{-0.013}$ & $1.170 \pm 0.082 \,^{+0.018}_{-0.058}$\\
1.0-1.2 & $0.818 \pm 0.034 \,^{+0.031}_{-0.009}$ & $0.851 \pm 0.033 \,^{+0.010}_{-0.010}$ & $0.781 \pm 0.031 \,^{+0.009}_{-0.009}$\\
1.2-1.4 & $0.648 \pm 0.026 \,^{+0.016}_{-0.008}$ & $0.693 \pm 0.026 \,^{+0.009}_{-0.008}$ & $0.661 \pm 0.023 \,^{+0.011}_{-0.008}$\\
1.4-1.6 & $0.484 \pm 0.019 \,^{+0.013}_{-0.006}$ & $0.499 \pm 0.018 \,^{+0.009}_{-0.007}$ & $0.470 \pm 0.017 \,^{+0.013}_{-0.006}$\\
1.6-1.8 & $0.408 \pm 0.016 \,^{+0.008}_{-0.007}$ & $0.382 \pm 0.015 \,^{+0.008}_{-0.006}$ & $0.348 \pm 0.013 \,^{+0.009}_{-0.005}$\\
1.8-2.0 & $0.320 \pm 0.014 \,^{+0.006}_{-0.007}$ & $0.308 \pm 0.008 \,^{+0.009}_{-0.006}$ & $0.255 \pm 0.010 \,^{+0.009}_{-0.004}$\\
2.0-2.4 & $0.206 \pm 0.006 \,^{+0.004}_{-0.004}$ & $0.194 \pm 0.006 \,^{+0.006}_{-0.003}$ & $0.169 \pm 0.005 \,^{+0.005}_{-0.003}$\\
2.4-2.8 & $0.109 \pm 0.004 \,^{+0.003}_{-0.002}$ & $0.106 \pm 0.004 \,^{+0.003}_{-0.002}$ & $0.106 \pm 0.004 \,^{+0.005}_{-0.002}$\\
2.8-3.2 & $0.065 \pm 0.003 \,^{+0.002}_{-0.002}$ & $0.057 \pm 0.003 \,^{+0.002}_{-0.001}$ & $0.053 \pm 0.003 \,^{+0.003}_{-0.001}$\\
3.2-4.0 & $0.031 \pm 0.001 \,^{+0.001}_{-0.001}$ & $0.029 \pm 0.001 \,^{+0.001}_{-0.001}$ & $0.025 \pm 0.002 \,^{+0.001}_{-0.001}$\\
4.0-5.0 & $0.010 \pm 0.001 \,^{+0.001}_{-0.000}$ & $0.010 \pm 0.001 \,^{+0.000}_{-0.000}$ & $0.009 \pm 0.001 \,^{+0.000}_{-0.000}$\\
\end{tabular}
}
\vspace{-0.29cm}
}

\newcommand{\XSecTableBinnedcombinedC}{
\center
\XSectionProjTextSize
\small{
\begin{tabular}{cccc}
\hline
\pt / \y & 3.52-3.70 & 3.70-3.88 & 3.88-4.06\\\hline
0.6-0.8 & $1.341 \pm 0.158 \,^{+0.034}_{-0.207}$ & $1.164 \pm 0.157 \,^{+0.030}_{-0.065}$ & $1.341 \pm 0.193 \,^{+0.120}_{-0.036}$\\
0.8-1.0 & $0.816 \pm 0.075 \,^{+0.013}_{-0.035}$ & $1.065 \pm 0.075 \,^{+0.018}_{-0.059}$ & $0.975 \pm 0.115 \,^{+0.018}_{-0.070}$\\
1.0-1.2 & $0.785 \pm 0.032 \,^{+0.010}_{-0.012}$ & $0.690 \pm 0.031 \,^{+0.010}_{-0.011}$ & $0.760 \pm 0.039 \,^{+0.013}_{-0.039}$\\
1.2-1.4 & $0.609 \pm 0.023 \,^{+0.012}_{-0.008}$ & $0.561 \pm 0.022 \,^{+0.010}_{-0.008}$ & $0.531 \pm 0.027 \,^{+0.012}_{-0.010}$\\
1.4-1.6 & $0.484 \pm 0.018 \,^{+0.016}_{-0.007}$ & $0.433 \pm 0.017 \,^{+0.011}_{-0.007}$ & $0.409 \pm 0.021 \,^{+0.016}_{-0.008}$\\
1.6-1.8 & $0.336 \pm 0.013 \,^{+0.008}_{-0.006}$ & $0.315 \pm 0.014 \,^{+0.011}_{-0.006}$ & $0.279 \pm 0.014 \,^{+0.011}_{-0.006}$\\
1.8-2.0 & $0.231 \pm 0.010 \,^{+0.006}_{-0.004}$ & $0.228 \pm 0.011 \,^{+0.009}_{-0.005}$ & $0.213 \pm 0.011 \,^{+0.007}_{-0.005}$\\
2.0-2.4 & $0.164 \pm 0.005 \,^{+0.007}_{-0.003}$ & $0.140 \pm 0.005 \,^{+0.006}_{-0.002}$ & $0.131 \pm 0.006 \,^{+0.003}_{-0.003}$\\
2.4-2.8 & $0.082 \pm 0.002 \,^{+0.004}_{-0.002}$ & $0.078 \pm 0.004 \,^{+0.003}_{-0.002}$ & $0.070 \pm 0.004 \,^{+0.004}_{-0.002}$\\
2.8-3.2 & $0.059 \pm 0.003 \,^{+0.004}_{-0.002}$ & $0.049 \pm 0.003 \,^{+0.002}_{-0.001}$ & $0.039 \pm 0.003 \,^{+0.006}_{-0.001}$\\
3.2-4.0 & $0.022 \pm 0.001 \,^{+0.001}_{-0.001}$ & $0.019 \pm 0.001 \,^{+0.002}_{-0.000}$ & $0.022 \pm 0.002 \,^{+0.003}_{-0.001}$\\
4.0-5.0 & $0.008 \pm 0.001 \,^{+0.001}_{-0.000}$ & $0.007 \pm 0.001 \,^{+0.001}_{-0.000}$ & $0.007 \pm 0.002 \,^{+0.000}_{-0.002}$\\
\hline\hline
\end{tabular}
}
}
\newcommand{\PtProjMeanLHCbData}{1.24 \pm 0.01 \gevc}
\newcommand{\PtProjMeanLHCbMC}{1.077 \gevc}
\newcommand{\PtProjMeanPerugiaMC}{1.238 \gevc}

\newcommand{\PtProjYSlopeLHCbData}{-44 \pm 27\mub }
\newcommand{\PtProjYSlopeLHCbMC}{-181 \pm 2\mub }
\newcommand{\PtProjYSlopePerugiaMC}{-149 \pm 3\mub }

\renewcommand{\XSectionProjTextSize}{\small}

\begin{document}
\begin{titlepage}
\pagenumbering{roman}

\vspace*{-1.5cm}
\centerline{\large EUROPEAN ORGANIZATION FOR NUCLEAR RESEARCH (CERN)}
\vspace*{1.5cm}
\hspace*{-0.5cm}
\begin{tabular*}{\linewidth}{lc@{\extracolsep{\fill}}r}
\ifthenelse{\boolean{pdflatex}}
{\vspace*{-2.7cm}\mbox{\!\!\!\includegraphics[width=.14\textwidth]{figs/lhcb-logo.pdf}} & &}%
{\vspace*{-1.2cm}\mbox{\!\!\!\includegraphics[width=.12\textwidth]{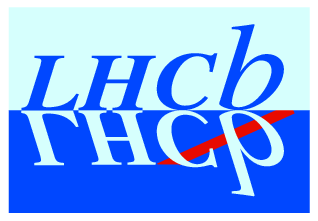}} & &}%
\\
 & & LHCb-PAPER-2011-007, CERN-PH-EP-2011-106 \\
 & & (submitted to PLB)\\
 & & \today \\
 & & \\
\end{tabular*}

\vspace*{4.0cm}

{\bf\boldmath\huge
\begin{center}
Measurement of the inclusive $\phi$ cross-section in \pp collisions at $\sqrt{s}=7\tev$
\end{center}
}

\vspace*{2.0cm}

\begin{center}
The LHCb Collaboration
\footnote{Authors are listed on the following pages.}
\end{center}

\vspace{\fill}

\begin{abstract}
  \noindent
The cross-section for inclusive $\phi$ meson production in \pp collisions at a centre-of-mass energy of $\sqrt{s}=7\tev$ has been measured with the LHCb detector at the Large Hadron Collider.
The differential cross-section is measured as a function of the $\phi$ transverse momentum \pt and rapidity \y in the region $\XSecRange$.
The cross-section for inclusive $\phi$ production in this kinematic range is $\XSecResultTotalcombined$, where the first systematic uncertainty depends on the \pt and \y region and the second is related to the overall scale. Predictions based on the \pythiasix generator underestimate the cross-section.
\end{abstract}

\vspace*{2.0cm}
\vspace{\fill}

\end{titlepage}

\newpage
\setcounter{page}{2}
\mbox{~}
\newpage

% forMemCo/LHCb-PAPER-2011-007/ on 18.07.2011

\begin{flushleft}
\section*{LHCb Collaboration}
% {\small today is 17. Jun. 2011}\\[4ex]
R.~Aaij$^{23}$, 
B.~Adeva$^{36}$, 
M.~Adinolfi$^{42}$, 
C.~Adrover$^{6}$, 
A.~Affolder$^{48}$, 
Z.~Ajaltouni$^{5}$, 
J.~Albrecht$^{37}$, 
F.~Alessio$^{6,37}$, 
M.~Alexander$^{47}$, 
G.~Alkhazov$^{29}$
% ...
\begin{footnotesize}
{\it
% ...
$ ^{5}$Clermont Universit\'{e}, Universit\'{e} Blaise Pascal, CNRS/IN2P3, LPC, Clermont-Ferrand, France\\
$ ^{6}$CPPM, Aix-Marseille Universit\'{e}, CNRS/IN2P3, Marseille, France\\
% ...
$ ^{23}$Nikhef National Institute for Subatomic Physics, Amsterdam, Netherlands\\
% ...
$ ^{29}$Petersburg Nuclear Physics Institute (PNPI), Gatchina, Russia\\
% ...
$ ^{36}$Universidad de Santiago de Compostela, Santiago de Compostela, Spain\\
$ ^{37}$European Organization for Nuclear Research (CERN), Geneva, Switzerland\\
% ...
$ ^{42}$H.H. Wills Physics Laboratory, University of Bristol, Bristol, United Kingdom\\
% ...
$ ^{47}$School of Physics and Astronomy, University of Glasgow, Glasgow, United Kingdom\\
$ ^{48}$Oliver Lodge Laboratory, University of Liverpool, Liverpool, United Kingdom\\
% ...
}
\end{footnotesize}

\end{flushleft}

\cleardoublepage

\pagestyle{plain}
\setcounter{page}{1}
\pagenumbering{arabic}

%==========================================
\section{Introduction}
Two specific regimes can be distinguished in hadron production in \pp collisions:
the so called hard regime at high transverse momenta, which can be described by perturbative QCD; and the soft regime, which is described by phenomenological models.
The underlying event in \pp processes falls into the second category. Therefore soft QCD interactions need careful study to enable tuning of the models at a new centre-of-mass energy.
Strangeness production is an important ingredient of this effort. 
Measurements of $\phi$ production have been reported by various experiments \cite{Anderson1976,Daum1981205,Alexopoulos:1995ru,HeraB2007,Chekanov:2002pt,Aamodt:2011zz,Naglis:2009uu} in different collision types, for different centre-of-mass energies and different kinematic coverage. 
LHCb is fully instrumented in the forward region and thus yields unique results complementary to previous experiments and to the other LHC experiments.

A measurement of the inclusive differential $\phi$ cross-section in \pp collisions at \mbox{$\sqrt{s}=7\tev$} is presented in this paper.
The analysis uses as kinematic variables the $\phi$ meson transverse momentum \pt and the rapidity $\y = \frac {1}{2}\ln \left[ (E+p_z) / (E-p_z) \right] $ measured in the \pp centre-of-mass system\footnote{The detector reference frame is a right handed coordinate system with $+z$ pointing downstream from the interaction point in the direction of the spectrometer and the $+y$ axis pointing upwards.}.
$\phi$ mesons are reconstructed using the $\Kp\Km$ decay mode and thus the selection relies strongly on LHCb's RICH (Ring Imaging Cherenkov) detectors for particle identification (PID) purposes.
Their performance is determined from data with a \tagNprobe approach.
The measured cross-section is compared to two different Monte Carlo (MC) predictions based on \pythiasix \cite{pythia64}.

%==========================================
\section{LHCb detector and data set}
\label{sec:dataset}
Designed for precise measurements of $B$ meson decays, the LHCb detector is a forward spectrometer with a polar angle coverage with respect to the beam line of approximately $15$ to $300\;\text{mrad}$ in the horizontal bending plane, and $15$ to $250\;\text{mrad}$ in the vertical non-bending plane. 
The tracking system consists of the Vertex Locator (VELO) surrounding the \pp interaction region, a tracking station upstream of the dipole magnet, and three tracking stations downstream of the magnet.    

Particles travelling from the interaction region to the downstream tracking stations are deflected by a dipole field of around $4\,\rm{Tm}$, whose polarity can be switched. 
For this study, roughly the same amount of data was taken with both magnet polarities.

The detector has a dedicated PID system that includes two Ring Imaging Cherenkov detectors. 
RICH1 is installed in front of the magnet and uses two radiators (Aerogel and $\text{C}_4\text{F}_{10}$), and RICH2 is installed beyond the magnet, with a $\text{CF}_4$ radiator.
Combining all radiators, the RICH system provides pion-kaon separation in a momentum range up to $100\gevc$.
Downstream of the tracking stations the detector has a calorimeter system, consisting of the Scintillating Pad Detector (SPD), a preshower, the electromagnetic and the hadronic calorimeter, and five muon stations.
Details of the LHCb detector can be found in Ref.~\cite{LHCBDETECTOR}.

The study described in this note is based on an integrated luminosity of \lumitot of \pp collisions collected in May 2010, where the instantaneous luminosity was low.

The trigger system consists of a hardware based first level trigger and a high level trigger (HLT) implemented in software.
The first level trigger was in pass-through mode, whereas at least one track, reconstructed with VELO information, was required to be found by the HLT.
On Monte Carlo simulated events, this trigger configuration is found to be 100\% efficient for reconstructed $\phi$ candidates.
However, to limit the acquisition rate, a prescaling was applied.

The luminosity was measured by two van der Meer scans \cite{vandermeer} and a novel method measuring the beam geometry with the VELO, as described in Ref.~\cite{lhcblumi_jpsi}.
Both methods rely on the measurement of the beam currents as well as the beam profile determination. 
Using these results, the absolute luminosity scale is determined, using the method described in Ref.~\cite{Aaij:2010nx}, with a 3.5\% uncertainty, dominated by the knowledge of the beam currents.
The instantaneous luminosity determination is then based on a continuous recording of the hit rate in the SPD, which has been normalized to the absolute luminosity scale.
The probability for multiple \pp collisions per bunch-crossing was negligibly low in the data taking period considered here.
\par

As the RICH detectors are calibrated separately for the two magnet polarities, the measurement is carried out separately for each sample before combining them for the final result.

Trigger and reconstruction efficiencies are determined using a sample of $1.25\cdot 10^{8}$ simulated minimum bias events.
These have been produced in the LHCb MC setting, which is based on a custom \pythia tune for the description of \pp collisions, while particle decays are generally handled by \evtgen \cite{evtgen}.
The total minimum bias cross-section in LHCb MC simulation is $91.05 \mbarn$, composed of the following \pythia process types: $48.80 \mbarn$ inelastic-non-diffractive,  $2 \times 6.84\mbarn$ single diffractive, $9.19\mbarn$ double diffractive and $19.28\mbarn$ elastic.
Details on the LHCb MC setting can be found in Ref.~\cite{Belyaev:1307917}.

%==========================================
\section{Data selection and efficiencies}
\label{sec:selection}

Two oppositely charged tracks, each of which are required to have hits in both the VELO and the main tracking system, are combined to form \phiKK candidates. 
The RICH system provides kaon-pion separation for reconstructed tracks, which is crucial for the inclusive $\phi$ production analysis.
As a first step, at least one kaon is required to pass a tight cut based on the RICH response during the selection. In a second step, both kaons have to pass this criterion.
The samples of $\phi$ candidates passing the cuts of the first and second steps are referred to as ``tag'' and ``probe'' samples, respectively.
They are used to measure the PID efficiency in the selection as explained below. 
The reconstructed \KK mass is required to be between 995\mevcc and 1045\mevcc in both samples.

No cut designed to discriminate prompt and non-prompt $\phi$ mesons is applied in the selection, so the measurement includes both.
However, due to the high minimum bias cross-section compared to charm or beauty production, the non-prompt contribution is small; in MC simulation it is found to be 1.6\%.

The region of interest 
$\XSecPtRange$ and $\XSecYRange$
is divided into 9 bins in \y and 12 bins in \pt. The differential cross-section per bin in \pt and \y is determined by the equation:
\begin{linenomath}
\begin{equation}
	\label{eq:crosssectionexplizit}
	\frac{\text{d}^2\sigma}{\text{d}y \,  \text{d}p_T} = \frac{1}{\Delta \y \, \Delta \pt}\cdot%
        \frac{N_{\mathrm{tag}}}%
	{\calL  \cdot \eff_{\rm reco}\cdot \eff_{\rm pid} \cdot \BR(\phi\to\Kp\Km) }, %
\end{equation}
\end{linenomath}
where $N_{\mathrm{tag}}$ is the number of reconstructed $\phi$ candidates in the tag sample, 
$\calL$ the integrated luminosity 
and $\BR(\phi\to\Kp\Km)=(49.2 \pm 0.6)\%$ the branching fraction taken from Ref.~\cite{PDG2008}.
The selection efficiency is split up into two parts in Eq.~\ref{eq:crosssectionexplizit}:
reconstruction $\eff_{\rm reco}$, including the geometrical acceptance, and the PID efficiency $\eff_{\rm pid}$.
Both efficiencies are a function of the \pt and \y values of the $\phi$ meson and thus determined separately for each bin.

In the centre of the kinematic region, the reconstruction efficiency is of the order of 65-70\%. It drops to 30-40\% with low transverse momenta and high or low rapidity values. The PID efficiency is above 95\% in the centre of the kinematic region and drops to 60-70\% at the edges of the considered kinematic region.

The reconstruction efficiency is determined from simulation.
To limit the MC dependence, the PID efficiency is determined from data using the \tagNprobe method: in the $\phi$ selection, at least one of the two kaons is required to pass the PID criterion.
The number of $\phi$ candidates passing this requirement is given by $N_\text{tag}$.
In a subsequent step, both kaons must pass the PID criterion. The number of $\phi$ candidates passing this step is given by $N_\text{probe}$.
The efficiency $\eff_{\rm pid}$ that at least one of the two kaons from a $\phi$ candidate fulfils the kaon PID requirement for each bin is thus given by:
\begin{linenomath}
\begin{equation}
  \label{eq:effstrip2}
  \eff_{\rm pid} = 1 - \left(\frac{N_{\rm tag} - N_{\rm probe}}{N_{\rm tag} + N_{\rm probe}}\right)^2.
\end{equation}
\end{linenomath}

This formula is valid only if the efficiencies that the two kaons satisfy the requirements are independent.
However, owing to the variation of the RICH efficiency with track multiplicity, correlations between the values of the discriminant variable of the RICH are observed and are accounted for in the systematic uncertainty.

%==========================================
\section{Signal extraction}
\label{sec:signalextraction}
Simultaneous maximum likelihood fits to the $\phi$ candidate mass distributions on the tag and the probe samples are performed in each bin of \pt and \y to extract the signal yields. The number of reconstructed candidates without PID requirements $N_{\mathrm{reco}}=N_{\mathrm{tag}}/\eff_{\rm pid}$ is a free parameter in the fit.
A Breit-Wigner distribution convolved with a Gaussian resolution function is used to describe the signal shape
\begin{equation}
f_\text{sig} = \frac{1}{\left(m-m_0\right)^2+\frac 14 \Gamma^2/c^4}\ \otimes\ \exp{\left(-\frac{1}{2}\frac{m'^2}{\sigma^2}\right)}
\end{equation}
while the background shape is described by
\begin{equation}
 f_\text{bkg} = 1-\exp{(c_1\cdot(m-c_2))}
\end{equation}
containing two free parameters.

\begin{figure}[Ht]
\centering
\includegraphics[width=0.45\textwidth]{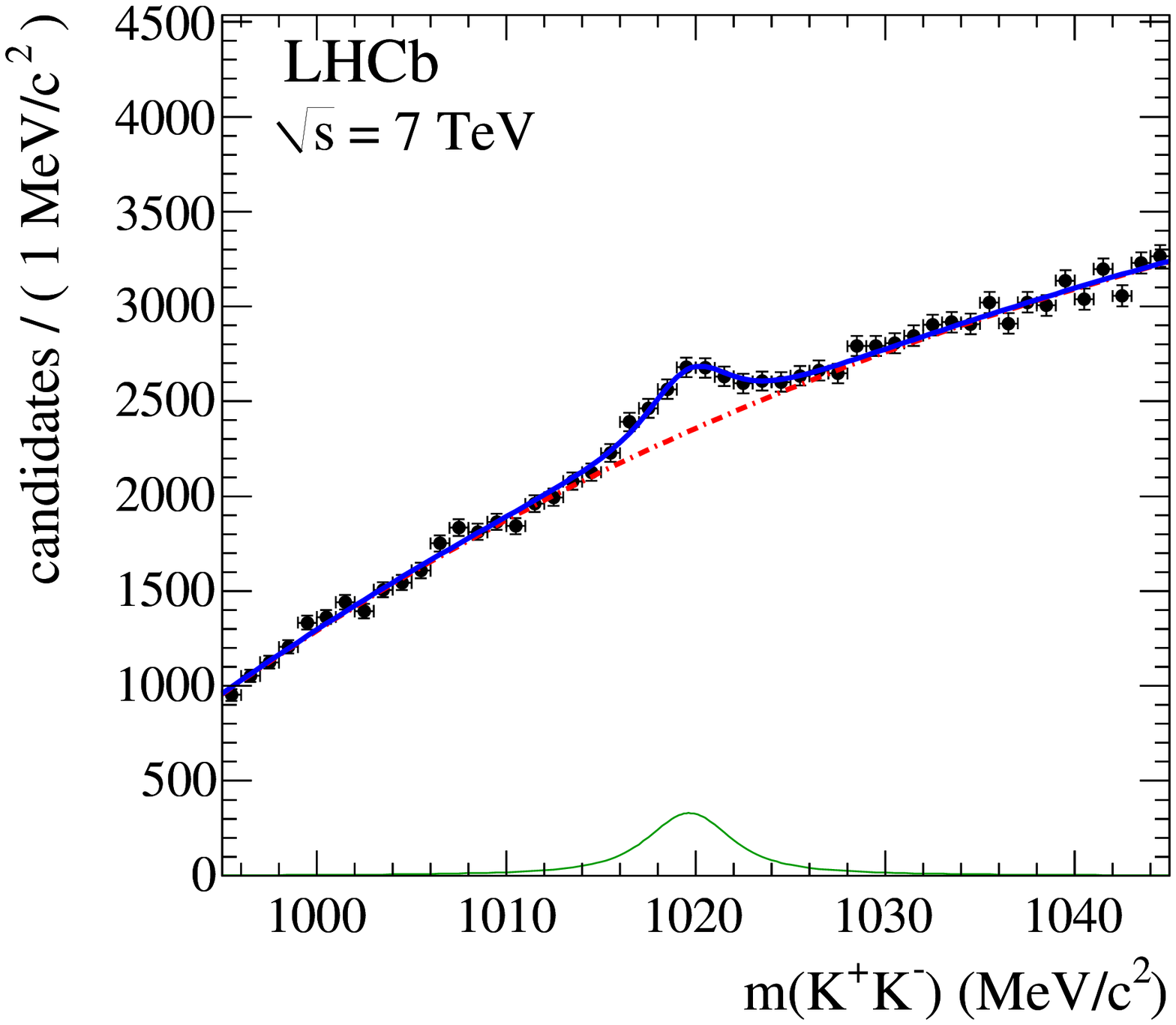}
\includegraphics[width=0.45\textwidth]{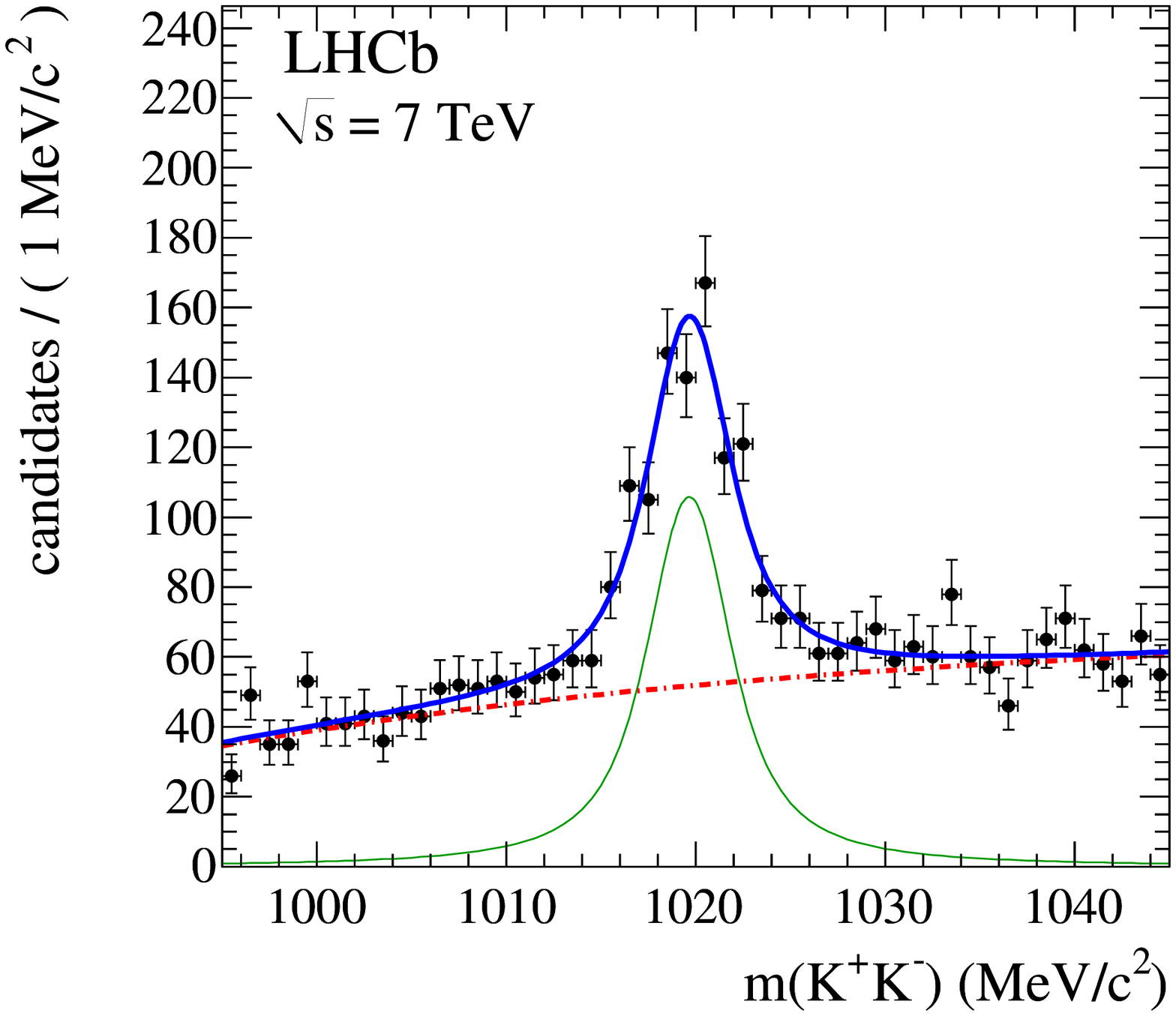}
\caption{Fit to the tag (left) and the probe (right) sample in the bin $0.6 < p_T < 0.8\gevc$, $3.34 < y < 3.52$ for one of the two magnet polarities. 
Shown are the data points, the fit result (thick solid line) as well as the signal (thin solid line) and the background component (dash-dotted line).
}
\label{fig:data_up_FIT_OR_pt4_y9}
\end{figure}

The fitted $\phi$ mass and the Gaussian width $\sigma$ are common parameters for both tag and probe sample, while the Breit-Wigner width $\Gamma$ is fixed to the value 4.26 \mev taken from Ref.~\cite{PDG2008}.
In Fig.~\ref{fig:data_up_FIT_OR_pt4_y9}, fit results to the two samples in a given \pt/\y bin are shown for illustration purposes.

%==========================================
\section{Systematic uncertainties}
\label{sec:systematics}
The uncertainties in this analysis are dominated by systematic contributions, divided into the ones which are common to all bins and the ones which vary from bin to bin.
The former are summarized in Table \ref{tab:syserrcorr}, whereas the latter are plotted with the data in Figure~\ref{fig:XSecProj} and listed in Table~\ref{tab:binnedresults}.
The bin-dependent uncertainties consist of the reconstruction efficiency uncertainty due to the limited simulation sample size and to the modelling of a diffractive contribution, as well as the uncertainty of the \tagNprobe PID determination due to correlations.
The combined uncertainties contribute 3--7\% for the statistically dominant bins.
\par

\begin{table}[!ht]
\caption{Summary of relative systematic uncertainties that are common to all bins.}
\center
\label{tab:syserrcorr}
\begin{small}
\begin{tabular}{lcrc}
\hline\hline
Source &  & (\%) & \\
\hline
Tracking efficiency         &  & 8 &\\
Luminosity (normalization)  &  & 4 &\\
Track multiplicity          &  & 3 &\\
Fit systematics             &  & 3 &\\
MC association              &  & 2 &\\
Doubly identified candidates&  & 2 &\\
Branching fraction          &  & 1 &\\
Bin migration               &  & 1 &\\
Material interactions       &  & 1 &\\
\hline
Total                       &  & 10 &\\
\hline\hline
\end{tabular}
\end{small}
\end{table}

The largest shared systematics are the uncertainty on the tracking efficiencies, which have been discussed in Ref.~\cite{Aaij2010209}, and the luminosity normalization.
The track multiplicity in data is higher than in the simulation.
Studies of the track multiplicity dependence of the reconstruction efficiency result in an uncertainty of 3\% due to this multiplicity difference.

Two major effects contribute to the uncertainty due to the fit procedure.
Fixing the Gaussian width to the same value on tag-and-probe sample introduces only a 1\% systematic uncertainty, since the distribution is dominated by the Breit-Wigner width.
A larger systematic effect (2-3\%) is observed when varying the mass range of the fit, which results in a total uncertainty of 3\%.

In the simulation, the reconstructed track is required to match the true generated track to determine the reconstruction efficiency. 
A 2\% uncertainty is assigned due to this procedure.
A small fraction of doubly identified candidates is found: it is possible that the detector hits from one particle are reconstructed as more than one track. The rate difference of these doubly identified candidates between data and simulation is found to be 2\%, which is the systematic uncertainty assigned due to this effect.
The \phiKK branching fraction contributes a 1\% systematic uncertainty.
Migration of candidates between different bins due to resolution effects is found to be small, and is accounted for by assigning a 1\% uncertainty.
Uncertainties from the modelling of the material budget and the material interaction cross-section are estimated to be 1\%.

%==========================================
\section{Results}
\label{sec:results}
The cross-sections determined with the two magnet polarities agree within their statistical uncertainties.
All results given here are unweighted averages of the two samples.
Comparisons to simulation samples generated with two different Pythia tunings are made, namely Perugia 0 \cite{SkandsPerugia} and the LHCb default Monte Carlo tuning.

\begin{figure}[Ht!]
\centering
\includegraphics[width=0.83\textwidth]{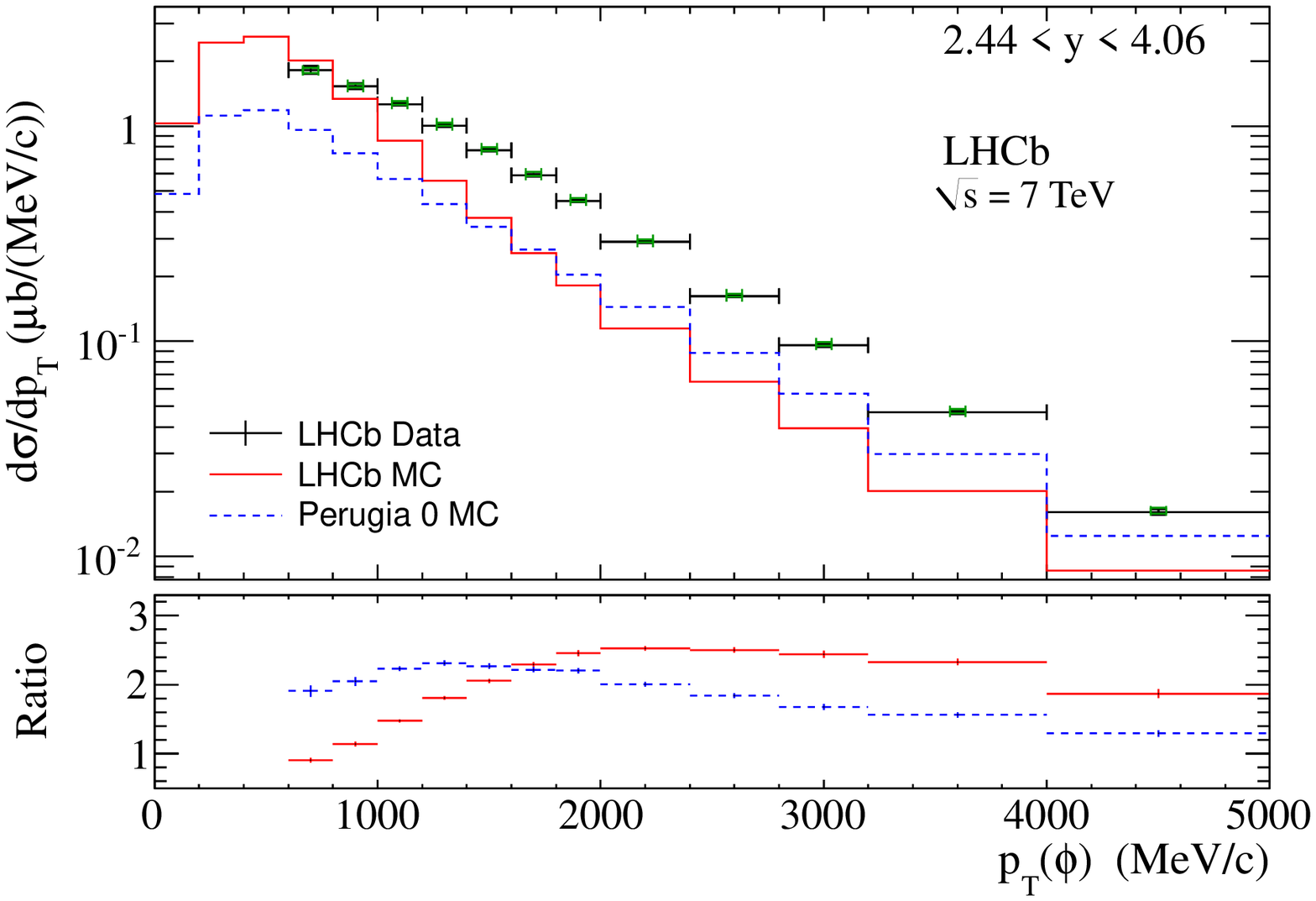}\\
\includegraphics[width=0.83\textwidth]{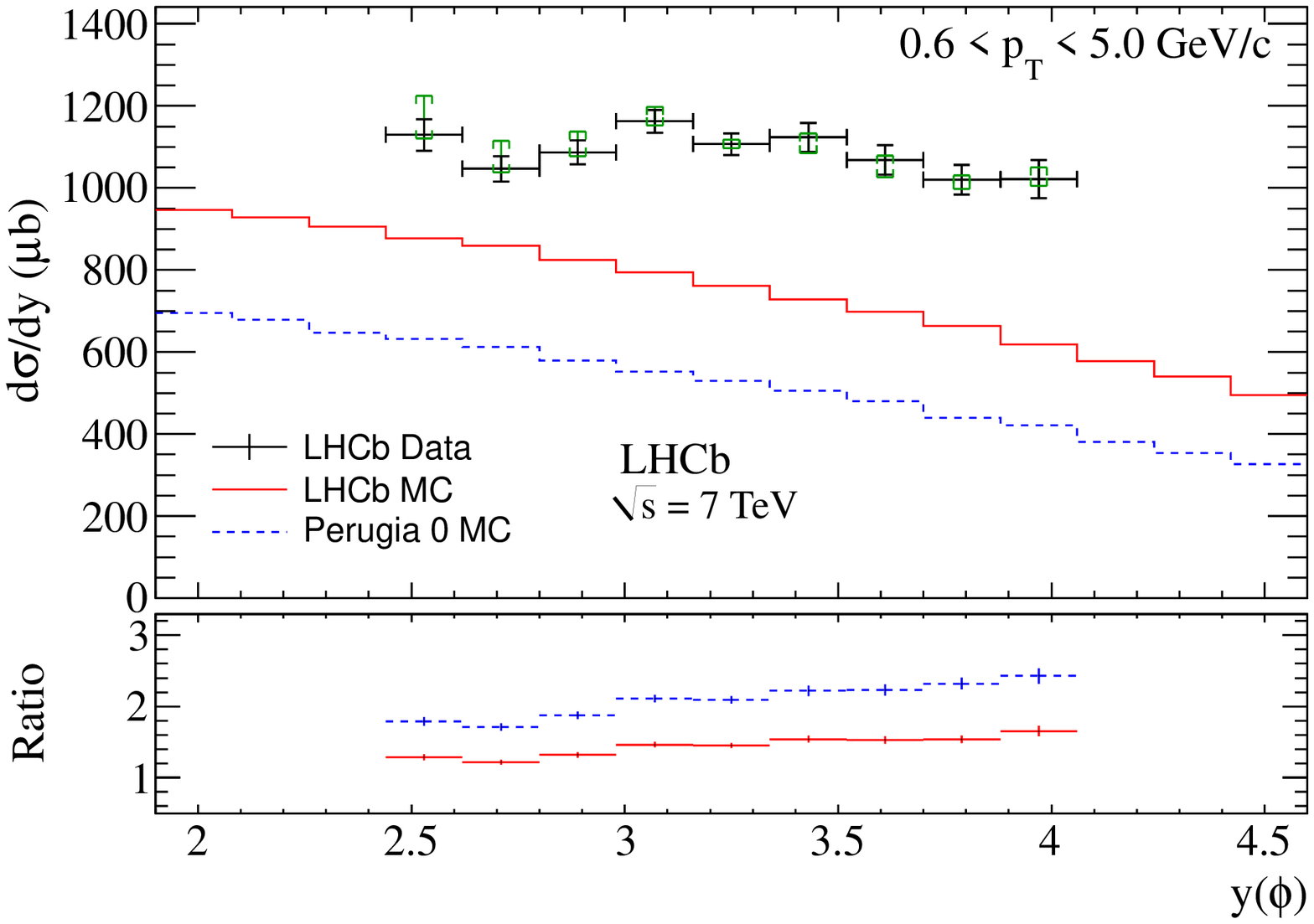}
\caption{Inclusive differential $\phi$ production cross-section as a function of
\pt (top) and \y (bottom), measured with data (points),
and compared to the LHCb default MC tuning (solid line) and Perugia 0 tuning (dashed line).
The error bars represent the statistical uncertainty, 
the braces show the bin dependent systematic errors, the overall scale uncertainty from Table~\ref{tab:syserrcorr} is not plotted. 
The lower parts of the plots show the ratio data cross-section over Monte Carlo cross-section.
Error bars in the ratio plots show statistical uncertainties only.}
\label{fig:XSecProj}
\end{figure}

The integrated cross-section in the region $\XSecRange$ is
\begin{linenomath}
$$\XSecResultTotalcombined,$$
\end{linenomath}

\noindent where the first systematic uncertainty arises from the bin-dependent contribution, while the second one is the common systematic uncertainty, as described in Section~\ref{sec:systematics}.
The differential cross-section values are given in Table~\ref{tab:binnedresults}
and projections on the \y and \pt axes within the same kinematic region are shown in Figure \ref{fig:XSecProj}.

The simulations underestimate the measured $\phi$ production in the considered kinematic region by a factor $\XSecResultTotalValRelLHCbMCcombined$ (LHCb MC) and $\XSecResultTotalValRelPerugiacombined$ (Perugia 0).
Additionally, the shape of the \pt spectrum and the slope in the \y spectrum differ between the data and the simulation (see Fig.\ \ref{fig:XSecProj}).
Fitting a straight line $\frac{\text{d}\sigma}{\text{d}\y} = a\cdot y +b$ to the \y spectrum, the slope is 
$a=\PtProjYSlopeLHCbData$ on data, but $a=\PtProjYSlopeLHCbMC$ for the default LHCb MC tuning and $a=\PtProjYSlopePerugiaMC$ for the Perugia 0 tuning. 
Uncertainties given on $a$ are statistical only.\par

The mean \pt in the range $\XSecPtRange$ is $\PtProjMeanLHCbData$ (data, stat. error only), $\PtProjMeanLHCbMC$ (LHCb MC) and $\PtProjMeanPerugiaMC$ (Perugia 0 MC).

\begin{table}
\label{tab:binnedresults}
\caption{Binned differential cross-section, in ${\mub}/\mevc$, as function of \pt (GeV/c) and \y. 
The statistical and the bin-dependent systematic uncertainties are quoted.
There is an additional bin-independent uncertainty of 10\% related to the normalization (Table~\ref{tab:syserrcorr}).}
\small
\XSecTableBinnedcombinedA
\XSecTableBinnedcombinedB
\XSecTableBinnedcombinedC
\end{table}

%==========================================
\section{Conclusions}
\label{sec:conclusion}
 
A study of inclusive $\phi$ production in \pp collisions at a centre-of-mass energy of $7\tev$ at the Large Hadron Collider is reported.
The differential cross-section as a function of \pt and \y measured in the range $\XSecRange$ is
\mbox{$\XSecResultTotalcombined$}, where the first systematic uncertainty depends on the \pt and \y scale and the second is related to the overall scale. 
Predictions based on the \pythiasix generator underestimate the cross-section.

\section*{Acknowledgements}
We express our gratitude to our colleagues in the CERN accelerator departments for the excellent performance of the LHC.
We thank the technical and administrative staff at CERN and at the LHCb institutes, and acknowledge support from the National Agencies: CAPES, CNPq, FAPERJ and FINEP (Brazil); CERN; NSFC (China); CNRS/IN2P3 (France); BMBF, DFG, HGF and MPG (Germany); SFI (Ireland); INFN (Italy); FOM and NWO (Netherlands); SCSR (Poland); ANCS (Romania); MinES of Russia and Rosatom (Russia); MICINN, XUNGAL and GENCAT (Spain); SNSF and SER (Switzerland); NAS Ukraine (Ukraine); STFC (United Kingdom); NSF (USA).
We also acknowledge the support received from the ERC under FP7 and the R\'{e}gion Auvergne.

\clearpage

\bibliographystyle{../LHCb}
\bibliography{../Bibliography}

\end{document}